\documentclass[preprint2]{aastex}

\shorttitle{Long Term Quasar Variability}
\shortauthors{De Vries et al.}

\begin{document}
\title{Long Term Variability of SDSS Quasars}

\author{W. H. de Vries, R. H. Becker}
\affil{University of California, One Shields Ave, Davis, CA 95616}
\affil{Lawrence Livermore National Laboratory, L-413, Livermore, CA 94550}
\email{devries1@llnl.gov}

\and 

\author{R. L. White}
\affil{Space Telescope Science Institute, 3700 San Martin Drive,
Baltimore, MD 21218}

\begin{abstract}

We use a sample of 3791 quasars from the Sloan Digital Sky Survey
(SDSS) Early Data Release (EDR), and compare their photometry to
historic plate material for the same set of quasars in order to study
their variability properties. The time base-line we attain this way
ranges from a few months to up to 50 years. In contrast to monitoring
programs, where relatively few quasars are photometrically measured
over shorter time periods, we utilize existing databases to extend
this base-line as much as possible, at the cost of sampling per
quasar. Our method, however, can easily be extended to much larger
samples. We construct variability Structure Functions and compare
these to the literature and model functions. From our modeling we
conclude that 1) quasars are more variable toward shorter wavelengths,
2) their variability is consistent with an exponentially decaying
light-curve with a typical time-scale of $\sim 2$ years, 3) these
outbursts occur on typical time-scales of $\sim 200$ years. With the
upcoming first data release of the SDSS, a much larger quasar sample
can be used to put these conclusions on a more secure footing.

\end{abstract}


\section{Introduction}

It has been known for a long time that quasars are variable, both on
short (days to weeks) and long (10's of years) timescales (e.g.,
Heckman 1976). There seems to be a consensus that the mechanisms
causing these variations are distinct: the large amplitude short
time-scale variations (mainly in BL Lac and OVV sources) are thought
to be due to relativistic beaming effects (e.g., Bregman 1990, Fan \&
Lin 2000, Vagnetti et al. 2003), whereas the long-term small amplitude
variations could possibly be due to accretion disk instabilities
(e.g., Rees 1984, Siemiginowska \& Elvis 1997, Kawaguchi et al. 1998),
bursts of supernovae close to the nucleus (e.g., Terlevich et
al. 1992, Cid Fernandes, Aretxaga, \& Terlevich 1996), or even
micro-lensing events by Galactic compact objects (e.g., Hawkins
1993). Disentangling the various proposed models from observations has
been particularly hard, partly because of the non-uniformity of the
observations, samples, and applied methods in the literature
(cf. Table~1 of Giveon et al. (1999) for a nice summary).

The classical approach tackling the variability problem is to monitor
a relatively small sample of quasars (up to a few 100) over long
periods of time, up to 20 years in cases. While this provides a well
(time) sampled database of light-curves, it does so for a select set
of objects. Given the wide range of correlations, anti-correlations,
or the lack thereof, found for particular parameters across different
data-sets and monitoring programs, sample selection is clearly an
issue. In this Paper, therefore, we used the largest public quasar
data-set that has both multi-epoch optical photometry and redshift
information.  Only by enlarging the sample enough we feel that results
are not going to be dominated by a few number of atypical objects.

We chose to use the Sloan Digital Sky Survey (SDSS) Early Data Release
(EDR, Stoughton et al. 2002) quasar list, cross-correlated with the
Second Generation Guide Star Catalog\footnote{The Guide Star
Catalogue-II is a joint project of the Space Telescope Science
Institute and the Osservatorio Astronomico di Torino.} (GSC2, McLean
et al. 1998) and the Palomar Optical Sky Survey (POSS, Reid et
al. 1991) imaging data. This resulted in a set of 3791 quasars, for
which we have on average 7 epoch measurements over 2 passbands
(cf. Sect.~\ref{calib}).  While our approach does not provide very
well sampled light-curves for individual quasars, or even accurate
photometry, is does have a few distinct advantages. First of all, the
sample is at least an order of magnitude larger than previous studies,
minimizing the impact of peculiar sources. Secondly, all our sample
quasars have redshift information, so we can correct for time-dilation
effects. The time-baseline over which we have photometric data is
$\sim50$ years, larger than previous work. But most importantly, the
sample can be easily enlarged by significant factors, which has the
potential to improve on the work presented here by quite a bit. The
SDSS collaboration is preparing to make its first public data release
soon.

The Paper is structured as follows: in Sect.~\ref{calib} we present
the careful calibration needed before one can use the variability
data-set. Section \ref{structfunc} will discuss the results of our
data analysis, its comparison to literature data and Monte Carlo
modeling. The final Section (\ref{optvar}) discusses the SDSS optical
properties of the most variable subset of our sample.

\section{Sample Selection and Calibration}\label{calib}

Our sample is based on the SDSS EDR quasar list, which was cross-
correlated with the GSC2 catalog. This resulted in a list of 3791
quasars (out of 3814 from the EDR list) with positional matches better
than 3\arcsec. The epoch base-line was extended by including POSS data
from the 1950's.  Not all of the 3791 quasars were retrieved from the
older data, so for a subset of the quasars (1347 objects) the maximum
base-line is limited to $\sim10$ years.  Due to the spectroscopic
requirements, our sample has an r-band limit of $\sim20.5$ magnitude
(the mean is $\overline{r}=18.94$), considerably fainter than for
instance work done on the Palomar Green (PG) quasars ($B < 16$, e.g.,
Giveon et al. 1999, Trevese \& Vagnetti 2002).  The quasar sample is
drawn from a $\sim348$ square degree region of sky around zero
declination (between $-1.5\arcdeg$ and $1.5\arcdeg$), and a small
patch of $\sim33$ square degrees around 17$^{\rm h}$20$^{\rm m}$ and
$+60\arcdeg$ declination.

The data for the GSC2 was provided to us by the Space Telescope
Science Institute, and the POSS data were taken from the USNO-A2
catalog server at Flagstaff. The more recent (and larger) USNO-B1
catalog (Monet et al. 2003) does not have a lot of 1950's era
photometry, whereas USNO-A2 contains uniquely data from that epoch.

\subsection{Passband Calibration}\label{passcalib}

\begin{figure*}[tb]
\epsscale{1.8}
\plotone{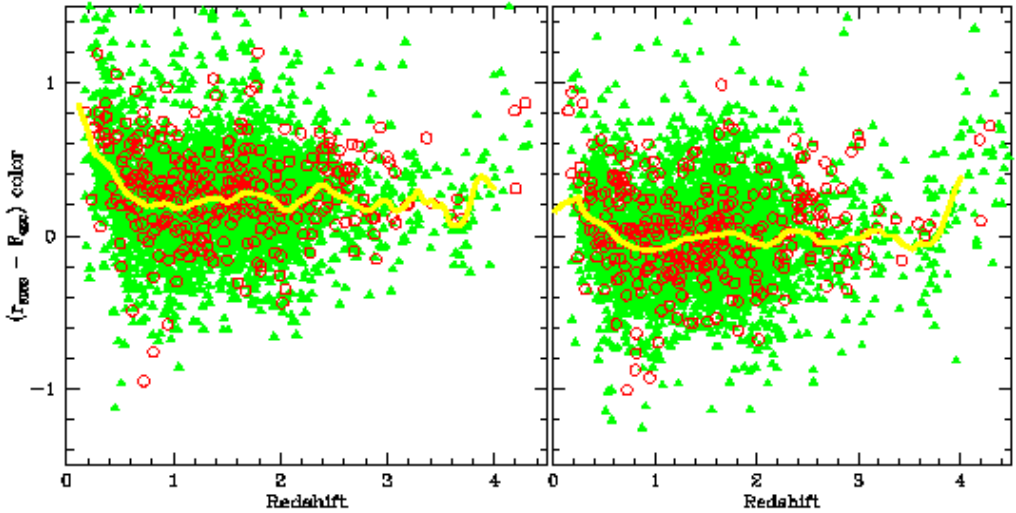}
\caption{These two panels illustrate the importance of proper passband
calibration. In the left panel, the uncorrected SDSS r-band to GSC2
F-band color is plotted as function of redshift. Note the offset
toward positive r$-$F colors, mainly due to the fact that SDSS
magnitudes are listed in the AB-mag system, and not in the Vega-mag
system. After properly transforming F-band into r-band magnitudes
using on average 36 stars per quasar, this residual color disappears
for the most part (right panel). The radio-quiet quasars are solid
green triangles, and the radio-loud quasars are open red circles,
respectively. The yellow solid line is the median color as function of
redshift. The upturn beyond $z=4$ is most likely due to absorption
shortward of Ly-$\alpha$ entering the r$_{sdss}$-band first.}
\label{passCalib}
\end{figure*}

Given that none of the photometric data have been taken with the same
telescope, detectors, and passbands, the relative calibration is
particularly important. Even though both the photographic USNO-A2 and
GSC2 catalogs have been calibrated with modern CCD measurements,
various uncontrollables, like plate material sensitivity changes /
degradation, render photographic plate calibration very difficult
(see, for instance, Gal et al. 2003). Instead of relying on the
cataloged magnitudes, we extracted all the stars within a 5 arcminute
radius around each quasar position. These stars can then be used to
calibrate and transform their magnitude (B and R for USNO-A2, and J
and F for the GSC2 catalog) into the corresponding SDSS g and
r-bands. The implicit assumption we make is that stars, on average,
are not variable. With mean numbers of calibration stars per quasar
of: 59 for USNO-A2, 47 for the GSC2, and 332 for the SDSS catalogs, we
feel confident we are not affected by a small number of variable
stars. All the extracted calibration stars were then position matched
to within 3 arcseconds of each other to make sure the same stars are
compared at the different epochs. This resulted in on average 36
position constant calibration stars per quasar (present in each
epoch), for a total of $\sim0.5$ million calibration points. For each
of the quasar fields, iterative least squares fits to the r$_{sdss}$
to R$_{usno-a2}$, r$_{sdss}$ to F$_{gsc2}$, g$_{sdss}$ to
B$_{usno-a2}$, and g$_{sdss}$ to J$_{gsc2}$\ color transformations
were made, fitting both the (linear) slope and offset.  To suppresses
the impact of stellar variability (and measurement errors) even
further, 2$\sigma$ outliers were excluded at each iteration.

The effectiveness of this calibration and passband transformation can
be seen in Fig.~\ref{passCalib}, where the uncorrected
(r$_{sdss}$$-$F$_{gsc2}$) color versus redshift distribution is
plotted in the left panel.  The thick solid line represents the sample
median (per redshift bin), illustrating the redshift independent color
offset for the quasars.  This apparent offset vanishes after applying
our stellar passband calibration (Fig.~\ref{passCalib}, right
panel). All that remains are the ``wiggles'' due to various emission
lines moving in and out of the passband (and which are very well
modeled / removed by synthetic passbands and a quasar composite
spectrum), and the apparent problem of measuring accurate magnitudes
of extended sources on the POSS and GSC2 plates.  The automated
routines work well on point-sources, but they systematically
overestimate resolved object magnitudes, resulting in large, positive,
(r$_{sdss}$$-$F$_{gsc2}$) values. This is illustrated by the upturn in
the median curve toward low redshifts. With the decrease in optical
size with increasing redshift, this effect becomes less important, and
is essentially absent beyond $z>0.6$.

While this calibration does take out plate-to-plate variations and
gross color terms, it does {\it not} remove slight offsets due to the
intrinsically different spectral shapes of quasars and stars. Fitting
the best color transformation using the stars actually ensures that
the color distribution (for the stars) has minimal scatter around a
mean of zero, which is exactly the presumption we made. This
minimization, while optimal for the stars, is not necessarily the best
for the quasars. The B$_{usno-a2}$ and J$_{gsc2}$
filters\footnote{effective photographic bandpasses, not physical
filters} have bandpasses {\it bluer} than g$_{sdss}$, whereas both the
R$_{usno-a2}$ and F$_{gsc2}$ bands are {\it redder} than the
corresponding r$_{sdss}$ (references for the photometric systems:
\cite{reid91} for POSS and GSC2; and \cite{stoughton02} for the SDSS).
Clearly, differences in spectral slopes for quasars and field stars
will give rise to offsets in the color distributions, and of opposing
signs between the blue and red bands.

\begin{figure}[tb]
\epsscale{1.05}
\plotone{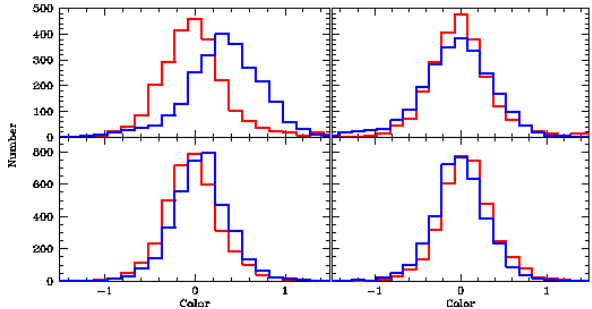}
\caption{Color distributions before (left) and after the
quasar--stellar spectral color correction. The top half of the plots
concern the (g$_{sdss}$$-$B$_{usno-a2}$) and
(r$_{sdss}$$-$R$_{usno-a2}$) colors (blue and red lines respectively),
whereas the bottom parts are the (g$_{sdss}$$-$J$_{gsc2}$), and
(r$_{sdss}$$-$F$_{gsc2}$) colors (same coloring). The applied offsets,
as determined by fitting Gaussian distributions to the histograms,
are: $+0.38$, $-0.02$, $+0.08$, and $-0.10$ (in the same color
order).}
\label{passCalib2}
\end{figure}

This can be directly tested by plotting up the relevant histograms, as
has been done in the left panel of Fig.~\ref{passCalib2}. If we assume
that the variations are random and distributed in a Gaussian fashion
(the histograms in Fig.~\ref{passCalib2} are actually well fitted by
Gaussians), we can measure the centroid offset accurately.  The best
fitting Gaussian functions, using a least squares (simulated
annealing) minimization code, have means of: $+0.38$, $-0.02$,
$+0.08$, and $-0.10$, for the color distributions
(g$_{sdss}$$-$B$_{usno-a2}$), (r$_{sdss}$$-$R$_{usno-a2}$),
(g$_{sdss}$$-$J$_{gsc2}$), and (r$_{sdss}$$-$F$_{gsc2}$)
respectively. The first indication that we are on the right track is
that the offset {\it signs} behave in the expected way: the same on
the blue (B and J to g) and red (R and F to r) side individually, and
opposite across the blue and red side for each epoch. We can, however,
do better. Assuming further that, since we are looking at mean
properties, we can use mean quasar and stellar spectra (folded with
the relevant passbands) to calculate what the offset values are
supposed to be.

\begin{figure}[tb]
\plotone{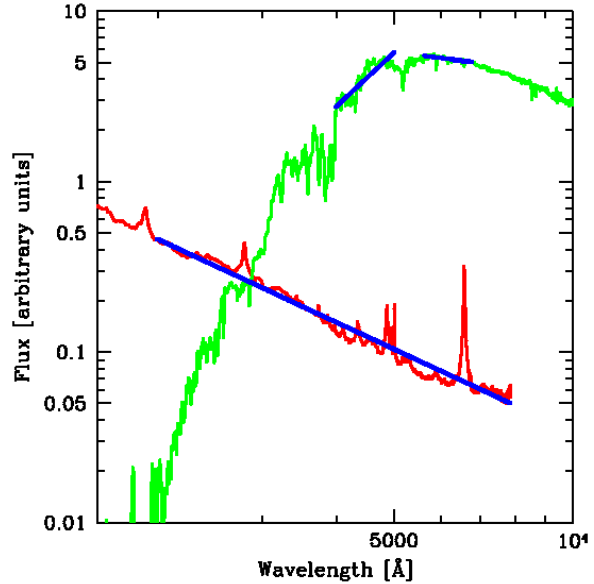}
\caption{Composite spectra of our quasars (red curve), and K2 main
sequence stars (green curve). The straight blue lines indicate the
fitted spectral slopes: $-1.62$ for the quasar, and $+3.25$ and
$-0.40$ for the stellar blue and red passbands. Note that these
slopes should be multiplied by 0.4 if put on the magnitude scale.}
\label{composite}
\end{figure}

We present both the quasars composite spectrum, and a mean-sequence
stellar spectrum (taken from the HILIB stellar flux library, Pickles
1998) in Fig.~\ref{composite}. Since the stars are always observed at
zero redshift, and the quasar slope is not depending on source
redshift (at least up to $z\approx 3$ for the blue), redshift effects
on the color terms can be ignored. The best fit to the measured values
is with a K2 stellar template, but even for a mean field star as blue
as G8, or red as K7, we get decent fits, cf. Table~1. A mean stellar
type of K2 is reasonable, given the combination of their relative
frequency and observed brightnesses of the calibration stars (between
R = 12 and R = 20), and is consistent with, for instance, the Bahcall
\& Soneira model for our Galaxy \citep{bahcall80,bahcall81}.  We do
note, however, that the calculated color offsets between the stars and
quasars are minimal (though still not zero for
g$_{sdss}$$-$B$_{usno-a2}$) with a mean F6 spectrum.

Given the predicted color offset values, and the reasonable assumption
of a K type dwarf as our mean stellar classification, we feel
confident that the bulk of the offsets are accounted for by this
quasar--stellar color offset. There may be other effects at work
though, like the Malmquist bias.  This bias represents the fact that
for a population with a negative number density gradient (i.e., the
number of sources increases with decreasing brightness), on average
more sources cross a selection limit toward increasing brightness,
than the other way around. So variable sources tend to be closer to
their peak brightness at their moment of selection (in our case, the
SDSS epoch), which in turn means that the color distributions should
have slightly negative mean values. Unfortunately, this weak signature
is lost in the color offset signal, as it is clear that we have both
positive and negative mean offsets in our uncorrected colors. The
values listed in Table~1 (and the ones for our adopted mean spectrum
of K2) have too many intrinsic uncertainties, like unknown and
variable details of the passbands, to be able to determine their
values with precision better than $\sim$0.05 magnitudes. Therefore,
retrieving a magnitude for the Malmquist offset from these particular
plots (like Helfand et al. 2001) is not feasible (but see
Sect.~\ref{optvar}).

It is very important for our structure function analysis
(cf. Sect.~\ref{structfunc}) to have symmetric color distributions
with means around zero, so we removed the color offset (including
possible Malmquist offsets) by applying the fitted means, and not just
the modeled values from Table~1. For the remainder of the Paper these
constants should be considered subtracted from each quasar color
individually, resulting in the corrected distributions in
Fig.~\ref{passCalib2}, right panel.

\section{Structure Function} \label{structfunc}

Structure Functions (SF hereafter) provide a tool to investigate the
time-dependence of variations. It is not very sensitive to aliasing
problems due to discrete (and sometimes sparse) time sampling, which
makes it well suited to our particular data-set. See Hughes, Aller, \&
Aller (1992, and references therein) for an introduction to SFs. We
define the SF as:

\begin{equation}
S(\tau) = \left( \frac{1}{N(\tau)} \sum_{i<j}{[m(i)-m(j)]^2} \right)^{\frac{1}{2}}
\end{equation}

\noindent analogously to \citet{hawkins02}. The summation is made over
the $N(\tau)$ measurement pairs for which $t_i-t_j = \tau$.  In
practice one calculates all the possible (positive) time-lag
permutations, amounting to $n(n-1)/2$ difference measurements, with n
being the total number of measurements.  After rebinning these into
discrete time-lag bins, the SF for each bin is simply given by the rms
in the magnitude variations. Since one usually has a large number of
difference measurements, a subsequent binning of the bins provides an
indication on the error in the SF measurement. This is typically what
we have employed in this paper.

\subsection{Stellar Structure Function}

\begin{figure}[tb]
\epsscale{1.0}
\plotone{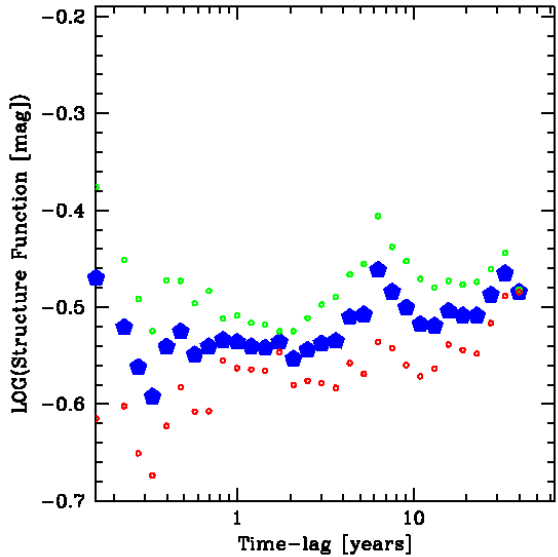}
\caption{Structure Function for our photometric calibration stars.  In
the absence of systematic increase of measurement uncertainty with age
of the observations, the SF should be constant. We detect a slight
increase toward longer time-lags, but both the rms level and increase
are small compared to our quasar SF. The blue pentagons represent the
combined data, and the green and red points above and below the
pentagons are the g-band and r-band data seperately.  Note that the
noise levels are lowest in the r-band. The peak in the green data
points around a time-lag of 7 years may be due to the relative paucity
of data at those time-lags.}
\label{StellarSF}
\end{figure}

There is some concern our data, since it has been taken at different
epochs with different equipment, may have a systematic bias toward
higher measurement noise with increasing age of the data. It is quite
clear that the most recent SDSS data have by far the best photometric
accuracy, and if for one reason or other, the 1950's era data would
happen to have a larger spread in their photometry compared to the
GSC2 data, we would have no trouble finding an increase in the rms as
a function of epoch. This would slant the SF upward with increasing
time-lag, away from the anticipated constant level. Any interpretation
based on such a SF would be suspect.

It is therefore that we set out to measure the SF for the calibration
stars separately. Again assuming our stars are not variable (on
average), their SF should be a constant, reflecting the overall mean
magnitude measurement uncertainty. One of the nice side-effects of the
quasar redshift distribution is that the time-sampling gets smeared
out from its initial measurement lumps around the 1950's, 1980's, and
2000, to a more or less uniform sampling by virtue of the $(1+z)$
time-frame correction. Since our calibration stars are all at $z=0$,
no such smoothing out is provided for. We therefore assigned a
``redshift'' to the individual stars to mimic this effect, randomly
selected from the quasar redshift distribution function\footnote{We
also assigned the calibration stars the redshift of their associated
quasar to test for possible hidden systematics. The resulting SF is
indistinguishable from the one presented here.}.

The resulting SF is plotted in Fig.~\ref{StellarSF}, for the combined
photometric data and the g- and r-bands separately. The plotting
ranges are identical to the other SF plots so it is easy to compare
them. While there is a slight increase in SF with increasing time-lag,
both the magnitude and slope do not affect our quasar SF to any level
of significance (cf. Fig.~\ref{SFN1} for example). We feel confident
that the trends in the quasar SF reflect real quasar variability
behavior.

\subsection{Structure Function Noise Dependencies}

\begin{figure}[tb]
\epsscale{1.0} 
\plotone{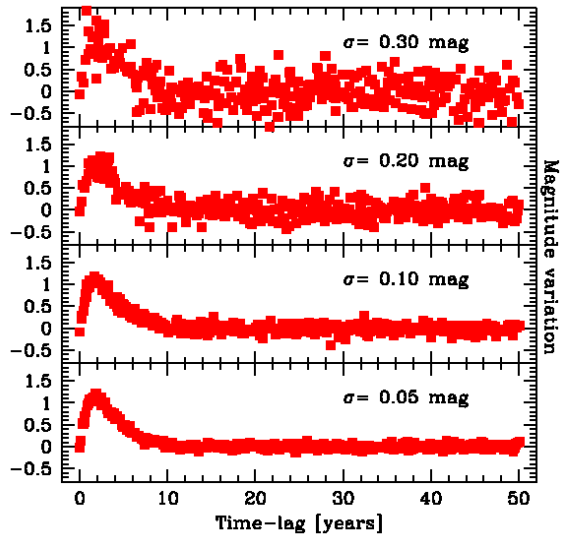}
\caption{Example light-curves with different levels of measurement
noise used to generate the structure functions in
Fig.~\ref{SFnoise}. The light-curve has the functional form given in
Eqn.~\ref{LCeqn}. The Gaussian noise (with indicated $\sigma$'s) has
been added to the randomly sampled function.}
\label{lightCurves}
\end{figure}

\begin{figure}[tb]
\epsscale{1.0}
\plotone{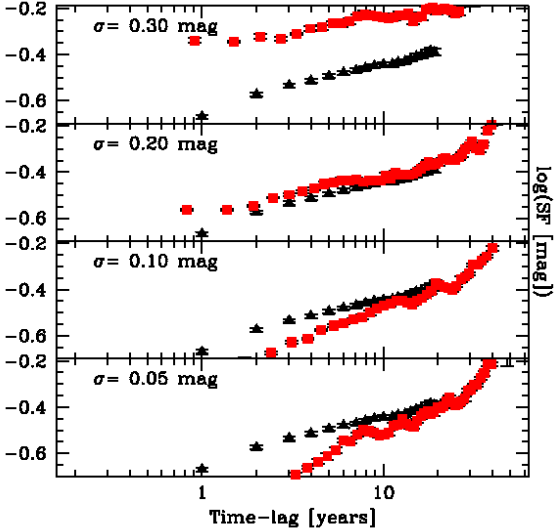}
\caption{Structure functions corresponding to the light-curves in
Fig.~\ref{lightCurves}. The largest noise-level dependence is seen for
short time-lags, where an increase in noise severely affects the SF
level and shape. At larger time-lags, the noise increase tends to
smooth out details in the SF, hiding both real and time-sampling /
aliasing signals. The overplotted data (small black triangles -
Hawkins 2002) would be adequately fitted by a $\sigma=0.15$ SF.}
\label{SFnoise}
\end{figure}

Analyses by \citet{kawaguchi98} and \citet{hawkins02} rely heavily on
the slope of the structure function. In the absence of noise (as noted
by Kawaguchi et al.), the slope can be correlated to different
light-curves, either due to star-formation, accretion disk
instabilities, or micro-lensing events. Since we know that our data
have significant ($\sigma \approx 0.2$ mag) photometric measurement
uncertainties, we need to know the effects of (white) noise on the
measurements. For this purpose, we calculated structure functions from
a particular light-curve, each with varying degrees of measurement
noise (cf. Fig.~\ref{lightCurves}). Aliasing effects due to a fixed
time sampling, which would affect the shorter time-lags
preferentially, have been minimized by randomly sampling the
light-curve. The noise, as added to the analytic form of the
light-curve, has a Gaussian distribution with a mean of 0, and the
$\sigma$ as indicated in the plots in Figs.~\ref{lightCurves} and
\ref{SFnoise}.

The light-curves presented in Fig.~\ref{lightCurves} are given by the
analytical expression (which prescribes both the rise and decay of an
outburst, and reaches its peak at $t=T$):

\begin{equation}
L(t,T)= A \left(\frac{e}{T}\right) t e^{-t/T}
\label{LCeqn}
\end{equation}

\noindent with $t$ in years, and $L(t,T)$ in magnitudes. The values of
$A$ and $T$, which represent the peak variation and brightness decay
timescale, have been set to 1.1 (mag) and 1.7 (years) respectively. We
chose this particular exponential declining light-curve since it
appeared to fit our data particularly well (see Sect.~\ref{qSF}). Even
though our data are actually very sparsely sampled light-curve
measurements of individual quasars (on average 3 measurements per band
per quasar), the large number of quasars in the sample do allow for an
adequate statistical sampling.

Figure \ref{SFnoise} depicts the effects of random (Gaussian)
photometric measurement noise on the resulting SF. Each light-curve
has been quasi randomly sampled in time (with $0 < \Delta t < 0.3$
years) to minimize aliasing effects in the SF due to a fixed sampling.
Our quasar data, once corrected for the source redshifts, exhibits a
fairly uniformly spaced light-curve sampling, with on the order of 5
measurements per year over the 50 year base-line. The model
light-curve is sampled at a comparable rate of $\sim 7$ measurements
per year. The four panels in the plot have increased noise levels in
the light-curves, starting at 0.05 mag in the bottom panel to 0.30
magnitudes in the top panel. A thing to keep in mind is that a {\it
uniform} noise level of 0.05, 0.10, 0.20, and 0.30 magnitudes
translates into a {\it constant} SF with values of $-$1.15, $-$0.85,
$-$0.55, and $-$0.37 respectively. It is immediately obvious that this
addition of a constant to the SF impacts the curve at short time-lags
(where its values are comparable in magnitude) the most: the SF slope
decreases significantly, up to the point where it is essentially zero
in the top panel. A slightly more subtle effect is also present, in
that small (time-scale) structure will get smoothed out.

Given our data quality, both of these effects have to be considered in
SF analyses, and obviously limit the extent to which we can push the
interpretation.

\subsubsection{Measurement Noise Levels}\label{measureNoise}

We can use the quasar SF, as plotted in Fig.~\ref{SFNtotal}, to
estimate the intrinsic photometric measurement uncertainties in our
data. Since at short time-lags the SF will be dominated by the
measurement noise, we can read off the $\log(S)$ level of this
``plateau'' directly. This is related to the measurement noise (see,
e.g., Hughes et al. 1992) through:

\begin{equation}
\sigma_{\mbox{noise}} = \frac{1}{\sqrt{2}}
10^{\log(S_p)} \quad \mbox{[mag]}
\end{equation}

\noindent which for a value of $\log(S_p) \sim -0.58$ translates into
a photometric measurement noise of 0.18 magnitudes. So before any
model light-curve can be compared to our data, we have to add 0.18
magnitudes of noise.

\subsection{Quasar Structure Function} \label{qSF}

\begin{figure}[tb]
\epsscale{1.0}
\plotone{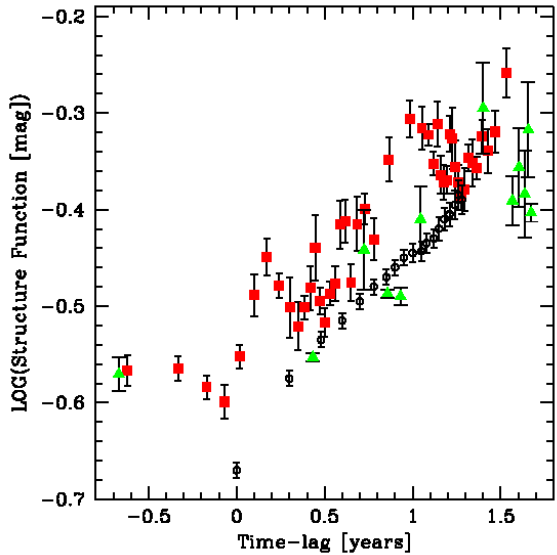}
\caption{Structure function for our quasar sample. Solid, red squares
denote time-dilation corrected variations (i.e. source intrinsic),
whereas the solid green triangles represent the SF as
observed. Overplotted as black circles are data from Hawkins (1996) on
a sample of 401 quasars, providing a better photometric quasar
baseline. Note the good agreement between the slopes of the data
sets. The Hawkins data can be best fitted to our (source intrinsic) SF
by offsetting it by 0.08 mag in the y-axis; it is already consistent
with the observed SF.}
\label{SFNtotal}
\end{figure}

\citet{kawaguchi98} correlated various SF slopes to different
intrinsic light-curves. For instance, their accretion disk instability
model results in typical SF slopes of 0.5, whereas models based on
variability due to starburst / supernovae have much steeper slopes
($\alpha \approx 0.9$). The relative robustness of these slopes with
respect to other model variables led \citet{hawkins02} to compare the SF
of a sample of 401 quasars with light-curve and redshift information
to these predicted model slopes. As another possible variability
mechanism, Hawkins suggested (micro)-lensing events, which produce an
SF slope $\alpha=0.25$ (cf. Table~1 of Hawkins). It is this slope that
is consistent with the Hawkins quasar data, suggesting a micro-lensing
origin of variability.

In Fig.~\ref{SFNtotal}, we present our SF and compare it to the
Hawkins data set. Since we have redshift information for all our
quasars, we can correct for time-dilation effects. Obviously, if the
variations are due to intervening micro lensing events, this
correction does not apply. The $(1+z)$ time-lag correction has the
nice side-effect of improving our time baseline sampling. Whereas the
non-corrected time-lags are necessarily quantized by the historic
observation dates, the corrected ones get spread out nicely. This
accounts for most of the difference between the two SF curves in
Fig.~\ref{SFNtotal}. The non-corrected data have gaps in their
time-lag coverage, resulting in a much less uniform data
distribution. The key information from this plot is that both
distributions have the same slope, and are consistent with the much
better time-sampled Hawkins data set. Effectively, the time-dilation
correction shifts the SF over toward shorter time-lags by 0.25 dex
along the x-axis. This translates into an inferred redshift of 0.78,
which is somewhat less than the median redshift of the sample
($z_{\mbox{med}}=1.46$, $z_{\mbox{ave}}=1.49$).

Unfortunately, unlike periodic photometric monitoring programs (e.g.,
the quasar light curve data base of Hawkins 1996) where each quasar is
observed multiple times and therefore provides input for each
particular time-lag bin (if sampled adequately), our data is much less
uniform. We can, however, combine both the g- and r-band datasets to
increase the time-sampling a bit, since the POSS and GSC2 surveys in
some cases did not take the green and red plates at the same
night. Even this only results in about 27\,000 magnitude permutations
which can be used to construct the SF. So, on average each quasar only
adds about 7 data points to the combined light curve. This also
implies that the set of quasars contributing to a particular time-lag
bin is not constant, which may be partly responsible for the larger
spread in the data (cf. Fig.~\ref{SFNtotal}) when compared to the
Hawkins (2002) data-set.

An added complication is that we do not have any idea about the actual
light-curve for each quasar. This does not mean that we cannot infer
anything about their shape however. If we assume that to first order
all quasar outbursts have comparable shapes (i.e. comparable half-life
timescales $T$), and occur on similar characteristic timescales $P$,
we can use the SF to constrain these parameters.  For this purpose we
created artificial SFs based on Monte-Carlo simulations of randomly
sampled light-curves. Each quasar gets assigned a randomly generated
light-curve with multiple outbursts of the form given in
Eqn.~\ref{LCeqn}. These outbursts are spaced in time with a
characteristic time-scale $P$. The probability of an outburst
occurring between $t$ and $t+dt$ is given by

\begin{equation}
\mbox{Prob($t$)} dt = \left(\frac{1}{P}\right) e^{-t/P} dt
\end{equation}

\noindent (which is identical to the Poisson distribution for the
probability that no event occurs in time $t$). This exponential
distribution is typical for shot noise models of variability, see
e.g., Lochner et al. (1991).

Since our model light-curves are noiseless, we added white noise (with
$\sigma=0.18$ mag, cf. Sect.~\ref{measureNoise}) to match the observed
noise characteristics. The resulting individual light-curves are more
similar to actual light-curves (e.g., Giveon et al. 1999) than the
ones presented in Fig.~\ref{lightCurves} for instance. From these
light-curves we then randomly selected 4 measurement points over a
$50/(1+z)$ year time period\footnote{This is assuming the variations
are intrinsic. For external variations (e.g., lensing), one has to use
the observed 50 year period.}, resulting in 6 time-lag permutations
which are entered into the database. This is done for all 3791
quasars, resulting in an artificial time-lag data-set which is
otherwise identical to the actual one, save for the fact that we know
it depends on the typical variation amplitude $A$, decay time $T$, and
characteristic outburst time-scale $P$.

To get a feel for how particular sets of these randomly sampled
light-curves translate into SFs, we note the following: 1) if $P$ is
small (on the order of 50 years or less), the SF shape is determined
by the signal due to the typical half-life time $T$. But since the
outbursts occur fairly regularly, there is a noticeable lack of
``variability contrast'' at time-scales close to (and beyond) $P$: the
SF flattens out. 2) the SF at short time-scales is controlled by the
half-life time $T$. For a given $T$, there is basically no real SF
signal at time-lags shorter than $T$.  3) if $P$ is large (on the
order of a few 100 years), and $T$ is relatively short ($<10$ years)
the outburst quickly become ``too rare'' on the large $P$ time-scale,
and noise starts dominating toward longer time-lags.  4) very short
$T$'s ($\sim1$ year) are progressively less time resolved given our
time-sampling and noise properties. This lessens our ability to make
any distinction between their cases.

We can attain acceptable fits to the SF for $T$ values ranging from 1
to 3 years, and $P$ values between 100 and 300 years. The ``best''
values are $T \approx 2$, $P \approx 200$ years, and $A \approx
2.1$. The value of $T$ is determined by the presence of signal in the
observed SF at time-lags of 1 to 3 years; modeled SFs with larger
values of $T$ do not have enough power at short time-lags to match the
observed SF, irrespective of the value of $P$.  The upper limit on $P$
is basically set by our noise limits, and therefore a significant
increase in sample size will result in a much harder limit. Lower
values of $P$ (at about 50 years) are ruled out since the SF starts to
lose power at comparable time-lags; it starts to flatten out beyond
what is seen in the data (Fig.~\ref{SFNtotal}).  It should be noted
that a value of $P \sim 100$ years is of the same magnitude as the
sound-crossing time for an accretion disk of a $10^9$ M$_\odot$ black
hole (cf. Courvoisier \& Clavel 1991, Trevese \& Vagnetti 2002); any
variations in the accretion rate would even out across the disk on
these time-scales.  Also, the $T \approx 2$ year value is remarkably
consistent with time-scales found by e.g., Cid Fernandes et al. (1996
and 2000, $1.5-3.0$ years), Cristiani et al.  (1996, $\sim 2.4$
years), Trevese et al. (1994, $\sim 1$ year), using a variety of data
and models. Even though our simulations (and data quality) do not
allow for an accurate assessment of $T$ and $P$, it is reassuring that
our results are close. It also implies that our initial assumption of
the existence of characteristic values for $T$ and $P$ for our quasars
is valid.

For non-source intrinsic variations (e.g., lensing), fits to the SF
tend to be less constrained because of the less uniform sampling of
our actual SF. None of the observed time-lags get corrected by their
uncorrelated source redshift, and therefore tend to cluster around a
few, $\sim 15$, and $\sim 40$ years respectively. Modeling results are
comparable to the ones for the intrinsic case, though much less
constrained, with a slightly higher likelihood for the combination $T
\approx 3$, $P \approx 400$, and $A \approx 2.2$. This is consistent
with the log 0.25 offset between the SF distributions
(cf. Fig.~\ref{SFNtotal}), which translates into a factor of 1.78
toward longer periods. With a larger data-set it might be possible to
discriminate between the intrinsic and external sources of variation
based on the SF. This is not possible with the current data-set,
however. 

Finally, as a consistency check, we can calculate the number of
quasars expected to have a variation more than $5\sigma=0.90$
magnitudes, given their characteristic light-curve and sample
size. This is done in Sect.~\ref{optvar}, but first we will
investigate the possible source intrinsic properties that may relate
to the SF shape.

\subsubsection{Color Dependencies}

\begin{figure}[tb]
\epsscale{1.0} 
\plotone{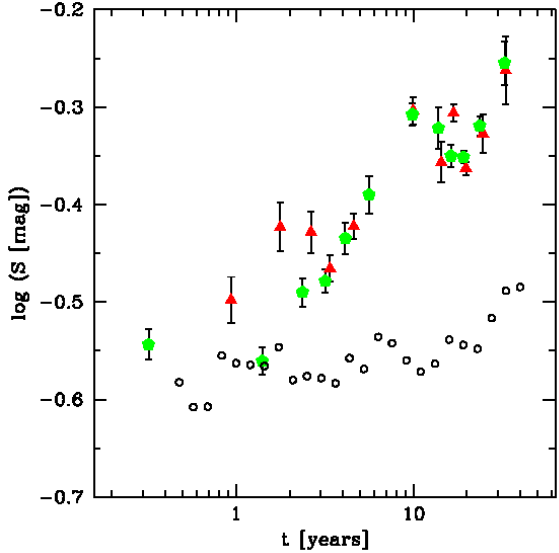}
\caption{Structure function for the quasars. The green pentagons and
red triangles represent the SF for the g- and r-bands respectively.
For time baselines beyond a few years, the green SF is consistently
more variable than the red one. The small black circles represent the
SF for the stars in the r-band, which provides a non-variable
base-line.}
\label{SFN1}
\end{figure}

It has been pretty well established that the magnitude of quasar
variation is a function of observed (and intrinsic) wavelength.
Giveon et al. (1999), for instance, noted a $0.02$ magnitude rms
difference between B- and R-band variability in a sample of 42 Palomar
Green (PG) quasars. The rms in the B$-$R color variations was found to
be $0.05$ mag. Variability shifts between the blue and red of this
magnitude have been found to be consistent over a range of samples
(cf. Trevese \& Vagnetti 2002). Wavelength variations in the near-IR
tend to be smaller, and are possibly too small to be measured (Enya et
al. 2002).

Possible mechanisms for these spectral variations include nuclear
star-bursts / supernovae which are predominantly blue (e.g., Aretxaga
et al. 1997, Cid Fernandes et al. 2000), and instabilities in the
nuclear accretion disk (e.g., Kawaguchi et al. 1998, Giveon et
al. 1999, Trevese et al. 2002). Even variations due to micro-lensing
(which are inherently non-intrinsic) could produce a wavelength
dependency if there is a {\it spatial} dependency on wavelength:
different parts will be amplified differently, giving rise to the
spectral variations. This would occur naturally for accretion disks
with a radial temperature gradient.

The two SFs for the g- and r-band are presented in Fig.~\ref{SFN1}.
Both SFs have the same slope, but are offset (vertically) by $0.03$ in
$\log(S)$, which is very close to the Giveon et al. (1999) value. This
translates into a 7\%\ (or $0.08$ mag) larger variation in $g$ than in
$r$, again numbers consistent with ones quoted from the
literature. The fact that both SFs are consistently offset beyond
$\sim 3$ year time-scales does not reflect any physical color
variability dependence (i.e. blue having different time-scales than
red), but more underlines the interdependence of the points in SF
plots. The slope is a far more robust discriminator (cf.  Kawaguchi et
al. 1998), and is identical for both bands.

\subsubsection{Source Intrinsic Dependencies}

Our sample contains about $8.5\%$ each of radio-loud and X-ray
luminous quasars (322 radio-loud, 328 X-ray luminous, with an overlap
of 52 quasars), so we can test whether these are more variable than
their radio-quiet and X-ray quiet counterparts. Evidence for this has
been found by e.g., Eggers et al. (2000). Furthermore, we can divide
the sample up in redshift bins to test for redshift (and consequently
luminosity, which is tightly coupled to the redshift)
dependencies. Unfortunately, this exercise cannot be carried too far,
as pretty quick systematic noise starts to dominate the signal of the
small number of quasars per bin.

The median redshift of our sample is $z=1.46$, so we divided our
sample up in a low- and high-redshift subset, with the median redshift
as discriminator. The resulting SFs are identical to the ones plotted
in Fig.~\ref{SFN1}, and no statistically significant offset is
seen. The only difference between the low- and high-redshift SFs is
that the latter does not extend toward long time-lags as much as the
low-redshift subset. This reflects the intrinsically smaller time-lags
for the high redshift bin compared to the low-redshift base-lines
(akin to the offset in Sect.~\ref{qSF}). Given the correlation between
redshift and (absolute) luminosity (cf. Fig.~\ref{hubbleQSR}), it does
not come as a surprise that we do not detect any significant
difference between the SFs of faint and bright quasars either. This is
further compounded by the fact that brighter quasars (which are on
average at higher redshifts) are found to be less variable (e.g., Hook
et al. 1994, Trevese et al. 1994), but are also sampled (given our
fixed bandpasses) progressively bluer, and more variable. These two
(small) effects therefore tend to cancel each other.  Improved
photometry and a larger sample may help in this respect.

The situation does not improve by isolating just the radio-loud and/or
X-ray luminous subsets of our sample. We cannot use the present
data-set to separate the radio-loud / X-ray loud SF from their quiet
counterpart.

\section{Optical Properties of Variable Quasars}\label{optvar}

\begin{figure}[tb]
\plotone{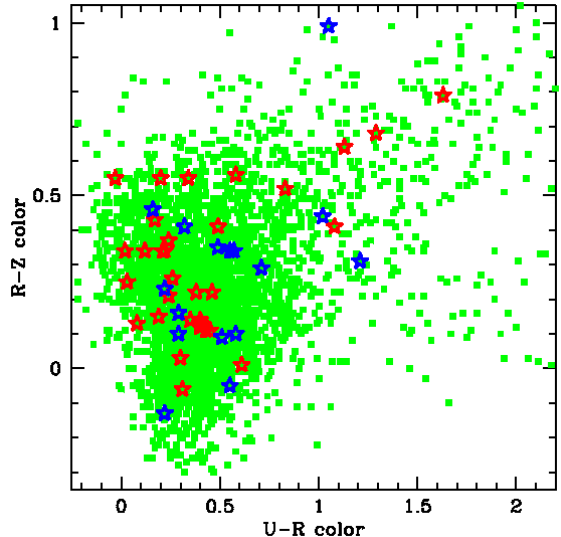}
\caption{SDSS Color-color plot of the most variable quasars (star
symbols) compared to the rest of the sample (small green squares).
The most variable quasars have been subdivided into ones that are
brighter (red stars - 34 objects) and fainter (blue stars - 17
objects) in the SDSS epoch. The disparity in the numbers is due to the
Malmquist bias. The small horizontal offset between the two groups may
be due to the spectral variability dependency.}
\label{mostVar}
\end{figure}

\begin{figure}[tb]
\plotone{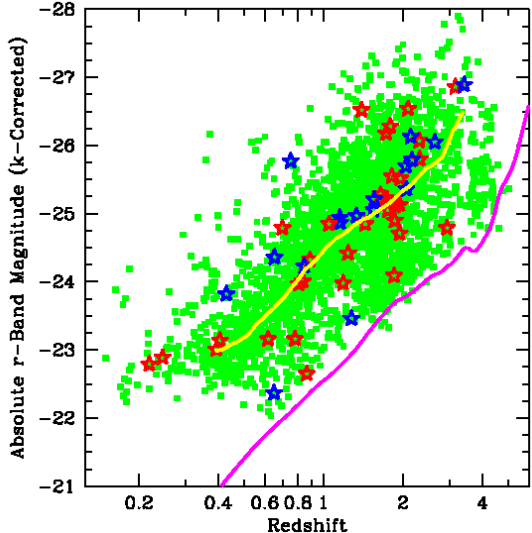}
\caption{Absolute r-Band magnitude of the quasar sample as function of
redshift (the ``Hubble Diagram''). The most variable quasars are
marked by the open star symbols (red for the ones that are brighter,
blue for the ones that are fainter in the SDSS epoch). Like in
Fig.~\ref{mostVar}, there is no significant difference between the
variable and non-variable populations.  All magnitudes have been
k-corrected using a quasar composite spectrum. The solid purple line
indicates the SDSS r-band detection limit of 20.5 magnitudes. The
yellow line represents the median absolute magnitude for the sample
(consistent with an observed $r=18.7$ median magnitude).  Both lines
are parallel, indicative of the usefulness of the composite spectrum.}
\label{hubbleQSR}
\end{figure}

From our initial quasar sample of 3791, we selected the 51 sources
with a g- and r-band rms in the variability of more than 0.9
magnitude, which corresponds to a variation of more than
$5\sigma$. These sources are tabulated in Table~2. This number of 51
sources is roughly consistent with the expectation based on a typical
light-curve with $T=2, P=200$ years. The length of time this
light-curve exhibits more than an 0.9 magnitude variation is about
$\sim 2.5\%$ of the time. This then linearly translates into $\sim 90$
or so quasars in our sample which would have a variability amplitude
of more than 5$\sigma$.

This most-variable (MV) subsample can be investigated in two
ways. First, the MV sources can be compared in its entirety to the
main set, and secondly, the MV source set can be intercompared with
itself by splitting it up in the ones that are brightest and the ones
that are faintest in the SDSS epoch (the epoch with the best
photometry).

Based on work by e.g., Trevese \& Vagnetti (2002), one expects a
color shift during outbursts. Since the quasars in the sub-set which
is brighter in the SDSS epoch are presumably undergoing an outburst
(in contrast to the fainter set), their colors should be offset from
the quasars in the fainter set. Unfortunately, we do not have the
non-outburst photometry for each particular quasar individually (at
least not to the required accuracy), so we have to fall back to
comparing the statistical properties of the samples.  As can be seen
in Fig.~\ref{mostVar}, the quasars span a large range in colors, and
simple average quantities will have large uncertainties associated
with them. This is further compounded by the very small number of
sources in each bin: 34 objects that are brighter in the SDSS epoch,
and 17 that are fainter by more than 5$\sigma$. This apparent
disparity between the numbers is due to the Malmquist Bias (as
discussed in Sec.~\ref{passcalib}).

We can minimize the impact of outlier points on average values by
applying proper weights. In this case we weight the points by their
distance from the simple means $\overline{u-r}$ and
$\overline{r-z}$. This results in weighted means of:
$\overline{u-r}_B=0.44$, $\overline{u-r}_F=0.65$,
$\overline{r-z}_B=0.28$, and $\overline{r-z}_F=0.28$, where the $B$
and $F$ indices indicate whether the sub-set was brighter or fainter
at the SDSS epoch. The weighted means for the complete quasar sample
are: $\overline{u-r}=0.52$ and $\overline{r-z}=0.23$. The color offset
between the faint and bright set in $\overline{u-r}$ (0.21 magnitudes)
is probably real, but their absolute value appears to be less certain,
given the fact that the overall sample weighted mean falls in between
the two. The effects in the $r-z$ colors are smaller due to the
smaller color range and the smaller color variability dependence.  A
spectral variability dependence, like B to R in the optical
\citep{trevese02} disappears (or becomes too small to measure) in the
near infrared \citep{enya02}.  Given the uncertainty in the average
numbers, studies of the spectral dependencies of variability are best
done on a small sample of quasars followed through their outburst,
instead of recovering the signal from a large sample of quasars.

Figure~\ref{hubbleQSR} illustrates the lack of absolute luminosity
dependence on variability. There is no evidence for less luminous
sources being more variable (as found by e.g., Hook et al. 1994), at
least not based on our data. \citet{cristiani96} find no correlation
between variability time-scales and absolute luminosity, which would
be more consistent with our findings. However, disentangling the
correlated effects of sample selection, detection limits, and
redshift-luminosity effects have been proven to be notoriously
difficult (cf. Giallongo, Trevese, \& Vagnetti 1991).

The main result of this section is to show that the (optical)
properties of quasars in outburst do not differ from their quiescent
phase, at least not in their average sense. The relative fraction of
highly variable quasars (both the bright and faint sets) is constant
over the range of absolute luminosities. The SDSS photometry should
be more than accurate enough to detect small deviations between the
two sub-sets, but revealed nothing beyond a small color dependency
signal, one that has been measured before \citep{trevese02}.

\section{Summary}

We have demonstrated the applicability of large photometric databases
in the long-term variability studies of quasars. The big advantage of
this method over detailed monitoring of a smaller set of sources is
its scalability. More sources, more catalogs, and more epochs can be
added to the existing data-set in a straightforward manner, all of
which will improve the robustness of the results. 

Our SF analysis of the initial set of 3791 quasars, as retrieved on
photographic plates dating from the 1950's through the 1980's,
combined with the SDSS era CCD photometry, yielded the following
results: 1) the variability as function of time-lag is totally
consistent with data-set based on close monitoring of quasars,
underlining the usefulness of the method; 2) quasars are more variable
toward shorter wavelengths, at least beyond time-lags of a few years;
3) their variability is consistent with an exponentially decaying
light-curve with a typical half-life time-scale of $\sim 2$ years; and
4) these outbursts occur on typical time-scales of $\sim 200$ years.

In a follow-up paper we will discuss the results of a much larger
sample of quasars, pending SDSS release of their first data-set.

\acknowledgments

It is a pleasure to thank Uriel Giveon and the referee, Dr. Trevese,
for useful comments that helped improve the paper.  WDVs work was
performed under the auspices of the U.S. Department of Energy,
National Nuclear Security Administration by the University of
California, Lawrence Livermore National Laboratory under contract
No. W-7405-Eng-48. This research has made use of the NASA/IPAC
extragalactic database (NED) which is operated by the Jet Propulsion
Laboratory, Caltech, under contract with the National Aeronautics and
Space Administration.

\begin{deluxetable}{lcccc}
\tablenum{1}
\tablecaption{Distribution Color Offsets Between Stars and Quasars\tablenotemark{1}}
\label{colorTab}
\tablehead{
  \colhead{Mean Stellar Template} &
  \colhead{(g$_{sdss}$$-$B$_{usno-a2}$)} & \colhead{(r$_{sdss}$$-$R$_{usno-a2}$)} &
  \colhead{(g$_{sdss}$$-$J$_{gsc2}$)} & \colhead{(r$_{sdss}$$-$F$_{gsc2}$)}
}
\startdata
Measured & $+0.38$ & $-0.02$ & $+0.08$ & $-0.10$ \\
\tableline
\\
F6V\tablenotemark{a}      & $+0.10$ & $-0.00$ & $+0.02$ & $-0.00$ \\
G0V      & $+0.15$ & $-0.01$ & $+0.03$ & $-0.02$ \\
G8V      & $+0.24$ & $-0.05$ & $+0.04$ & $-0.06$ \\
K0V      & $+0.24$ & $-0.04$ & $+0.04$ & $-0.08$ \\
K2V\tablenotemark{b}      & $+0.34$ & $-0.05$ & $+0.05$ & $-0.09$ \\
K7V      & $+0.54$ & $-0.10$ & $+0.06$ & $-0.21$ \\
M1V      & $+0.54$ & $-0.09$ & $+0.06$ & $-0.23$ \\
\enddata
\tablenotetext{1}{Units are in magnitudes, and are listed as stellar
mean $-$ quasar mean.}  
\tablenotetext{a}{Smallest overall color offset.}  
\tablenotetext{b}{Best fit to our observed offsets.}
\end{deluxetable}

\begin{deluxetable}{lrrrp{12mm}|lrrrrp{12mm}}
\tablenum{2}
\tablecaption{Most Variable Subset of Quasars}
\label{varTab}
\tablehead{
  \colhead{RA\tablenotemark{a}} & \colhead{DEC} & \colhead{z} & \colhead{r$_{sdss}$} & \colhead{Var\tablenotemark{b}}  & 
  \colhead{RA\tablenotemark{a}} & \colhead{DEC} & \colhead{z} & \colhead{r$_{sdss}$} & \colhead{Var\tablenotemark{b}}
}
\startdata
00 06 54.11 & $-$00 15 33.4 & 1.727 & 18.05 & $+$1.09  &  03 37 06.22 & $-$00 47 47.7 & 0.751 & 17.52 & $-$2.30\\
00 10 22.15 & $-$00 37 01.2 & 3.156 & 18.24 & $+$1.00  &  03 37 37.45 & $-$00 24 57.2 & 1.186 & 19.93 & $+$1.02\\
00 11 30.56 &  00 55 50.7 & 2.306 & 18.50 & $+$1.67    &  03 44 25.67 &  00 54 52.8 & 0.846 & 19.33 & $-$1.93\\
00 14 38.28 & $-$01 07 50.2 & 1.813 & 18.84 & $+$0.96\tablenotemark{c}&  10 22 24.82 &  00 06 42.5 & 0.617 & 19.36 & $+$1.09\tablenotemark{c}\\
00 16 02.40 & $-$00 12 25.0 & 2.090 & 17.90 & $+$1.23  &  10 23 16.61 &  00 09 36.3 & 0.836 & 18.98 & $+$1.45\\
00 16 57.00 &  00 55 32.0 & 1.754 & 19.19 & $+$0.94    &  10 28 36.87 & $-$00 43 14.3 & 0.219 & 17.86 & $+$1.08\\
00 17 35.69 & $-$01 13 25.1 & 0.805 & 18.99 & $+$1.17\tablenotemark{c}&  10 57 44.32 & $-$00 37 07.0 & 2.163 & 18.76 & $-$0.91\\
00 19 19.31 &  01 01 52.2 & 2.312 & 18.78 & $+$0.95    &  11 30 12.38 &  00 33 14.7 & 1.940 & 19.68 & $+$1.96\\
00 20 23.18 & $-$00 11 10.6 & 1.645 & 19.01 & $+$1.15  &  12 04 55.09 &  00 26 41.3 & 1.555 & 18.77 & $-$0.97\\
00 24 11.66 & $-$00 43 48.1 & 1.790 & 17.98 & $+$1.15\tablenotemark{c}&  13 06 11.19 & $-$00 09 32.9 & 0.393 & 19.19 & $+$1.06\\
00 27 23.44 & $-$00 16 13.2 & 1.708 & 18.92 & $+$1.37  &  13 07 24.75 &  00 37 38.5 & 0.652 & 18.12 & $-$1.05\\
00 45 16.00 &  00 00 42.3 & 1.548 & 18.88&  $-$2.33      &  13 11 08.48 &  00 31 51.8 & 0.429 & 17.93 & $-$1.13\\
01 21 00.73 & $-$00 15 19.0 & 0.864 & 20.35 & $+$0.93\tablenotemark{c,d}  & 13 22 56.51 & $-$00 59 30.2 & 1.152 & 18.46 & $-$2.57\\
01 41 36.39 & $-$00 10 19.6 & 0.405 & 18.48 & $+$1.73  &  13 38 56.85 &  00 27 18.3 & 1.954 & 18.87 & $+$1.38\\
01 42 14.74 &  00 23 24.2 & 3.405 & 18.27 &  $-$1.09      &  14 19 51.61 & $-$00 46 05.9 & 1.940 & 19.38 & $+$1.10\\
01 43 55.76 & $-$00 13 38.3 & 1.156 & 18.54 & $-$1.21    &  14 26 50.90 &  00 51 50.5 & 1.333 & 18.72 & $-$0.91\\
02 00 06.31 & $-$00 37 09.7 & 2.136 & 18.33 & $-$1.56    &  14 46 46.37 & $-$00 31 43.8 & 0.699 & 17.90 & $+$0.96\\
02 21 43.19 & $-$00 18 03.9 & 2.638 & 18.74 & $-$1.09    &  15 13 07.26 & $-$00 05 59.3 & 1.860 & 19.55 & $+$0.96\\
02 31 53.78 & $-$00 32 32.1 & 1.721 & 18.97 & $+$0.90  &  17 00 11.20 &  60 03 41.6 & 2.053 & 18.67 & $-$1.01\tablenotemark{d}\\
02 34 44.41 &  00 01 28.7 & 0.889 & 18.66 & $+$0.93    &  17 06 11.40 &  61 00 52.9 & 2.061 & 19.05 & $-$0.98\tablenotemark{d}\\
02 36 39.59 & $-$00 53 25.5 & 0.780 & 19.64 & $+$1.47  &  17 08 30.25 &  61 05 17.6 & 1.050 & 18.39 & $+$2.57\\
02 40 52.83 & $-$00 41 11.0 & 0.246 & 18.02 & $+$1.12  &  17 14 42.92 &  61 11 58.6 & 0.650 & 20.11 & $-$1.04\\
02 47 48.87 & $-$00 01 47.6 & 1.866 & 19.32 & $+$0.92  &  17 16 56.30 &  55 17 53.0 & 2.933 & 20.06 & $+$0.94\tablenotemark{c}\\
03 10 19.95 &  01 01 11.5 & 1.393 & 17.69 & $+$1.24    &  17 42 56.09 &  54 05 28.3 & 1.439 & 19.04 & $+$0.96\\
03 15 42.47 & $-$01 00 51.9 & 1.238 & 19.33 & $+$0.95  &  17 43 55.78 &  54 00 12.0 & 1.851 & 20.29 & $+$0.96\\
03 37 06.22 & $-$00 47 47.7 & 0.751 & 17.52 &$-$2.30    &  23 49 39.89 & $-$00 13 15.3 & 1.276 & 20.15 & $-$0.93\tablenotemark{c}\\
\enddata
\normalsize
\tablenotetext{a}{Units of right ascension are hours, minutes, and seconds, and units
of declination are degrees, arcminutes, and arcseconds. Epoch is J2000.}
\tablenotetext{b}{RMS of the optical variability in magnitudes. A '+' ('-') denotes quasars that
are brighter (fainter) in the SDSS epoch.}
\tablenotetext{c}{Detected in the radio by FIRST (1.4 GHz).}
\tablenotetext{d}{Detected in X-rays by ROSAT (0.5-2 KeV).}
\end{deluxetable}


\begin{thebibliography}{}

\bibitem[Aretxaga et al.(1997)]{aretxaga97} Aretxaga, I., Cid
Fernandes, R., \& Terlevich, R. J. 1997, \mnras, 286, 271

\bibitem[Bahcall \& Soneira(1980)]{bahcall80} Bahcall, J. N., \&
Soneira, R. M. 1980, \apjs, 44, 73

\bibitem[Bahcall \& Soneira(1981)]{bahcall81} Bahcall, J. N., \&
Soneira, R. M. 1981, \apj, 246, 122


\bibitem[Bregman et al.(1990)]{bregman90} Bregman, J. N., Glassgold,
A. E., Huggins, P. J., et al. 1990, \apj, 352, 574

\bibitem[Cid Fernandes et al.(1996)]{cid96} Cid Fernandes, R., 
Aretxaga, I., \& Terlevich, R. 1996, \mnras, 282, 1191

\bibitem[Cid Fernandes et al.(2000)]{cid00} Cid Fernandes, R., Sodr\'e
Jr, L., \& Vieira da Silva Jr, L. 2000, \apj, 544, 123

\bibitem[Courvoisier \& Clavel(1991)]{courvoisier91} Courvoisier,
T. J.-L., \& Clavel, J. 1991, \aap, 248, 389

\bibitem[Cristiani et al.(1996)]{cristiani96} Cristiani, S., Trentini,
S., La Franca, F., Aretxaga, I., Andreani, P., Vio, R., \& Gemmo, A.
1996, \aap, 306, 395

\bibitem[Eggers et al.(2000)]{eggers00} Eggers, D., Shaffer, D. B., \&
Weistrop, D. 2000, \aj, 119, 460

\bibitem[Enya et al.(2002)]{enya02} Enya, K., Yoshii, Y., Kobayashi,
Y., Minezaki, T., Suganuma, M., Tomita, H., \& Peterson, B. A. 2002,
\apjs, 141, 31

\bibitem[Fan \& Lin(2000)]{fan00} Fan, J. H., \& Lin, R. G., 2000,
\apj, 537, 101

\bibitem[Gal et al.(2003)]{gal03} Gal, R. R., de Carvalho, R. R.,
Odewahn, S. C., Djorgovski, S. G., Mahabal, A., Brunner, R. J., \&
Lopes, P. 2003, \aj, in press (astro-ph/0210298)

\bibitem[Giallongo et al.(1991)]{giallongo91} Giallongo, E.,
Trevese, D., \& Vagnetti, F. 1991, \apj, 377, 345

\bibitem[Giveon et al.(1999)]{giveon99} Giveon, U, Maoz, D., Kaspi,
S., Netzer, H., \& Smith, P. S. 1999, \mnras, 306, 637

\bibitem[Hawkins(1993)]{hawkins93} Hawkins, M. R. S. 1993, Nature,
366, 424

\bibitem[Hawkins(1996)]{hawkins96} Hawkins, M. R. S. 1996, \mnras,
278, 787

\bibitem[Hawkins(2002)]{hawkins02} Hawkins, M. R. S. 2002, \mnras,
329, 76

\bibitem[Heckman(1976)]{heckman76} Heckman, T. M. 1976, \pasp, 88, 844

\bibitem[Helfand et al.(2001)]{helfand01} Helfand, D. J., Stone,
R. P. S., Willman, B., White, R. L., Becker, R. H., Price, T., Gregg,
M. D., \& McMahon, R. G. 2001, \apj, 121, 1872

\bibitem[Hook et al.(1994)]{hook94} Hook, I. M., McMahon, R. G.,
Boyle, B. J., \& Irwin, M. J. 1994, \mnras, 268, 305

\bibitem[Hughes et al.(1992)]{hughes92} Hughes, P. A., Aller, H. D.,
\& Aller, M. F. 1992, \apj, 396, 469

\bibitem[Kawaguchi et al.(1998)]{kawaguchi98} Kawaguchi, T.,
Mineshige, S., Unemara, M., \& Turner, E. L. 1998, \apj, 504, 671

\bibitem[Lochner et al.(1991)]{lochner91} Lochner, J. C., Swank,
J. H., \& Szymkowiak, A. E. 1991, \apj, 376, 295

\bibitem[McLean et al.(1998)]{mclean98} McLean, B. J., Hawkins, G.,
Spagna, A., Lattanzi, M., Lasker, B. M., Jenkner, H., \& White, R. L.
1998, ``The Second Guide Star Catalogue'', in ``New Horizons from
Multi-Wavelength Sky Surveys'', proceedings of IAU Symposium 179, eds.
B. McLean, D. Golombek, J. Hayes, \& H. Payne, p. 431

\bibitem[Monet et al.(2003)]{monet03} Monet, D. G., Levine, S. E.,
Canzian, B., et al. 2003, \aj, 125, 984

\bibitem[Pickles(1998)]{pickles98} Pickles, A. J. 1998, \pasp, 110,
863

\bibitem[Rees(1984)]{rees84} Rees, M. J. 1984, ARA\&A, 22, 471

\bibitem[Reid et al.(1991)]{reid91} Reid I. N., Brewer, C., Brucato,
R. J., et al. 1991, \pasp, 103, 661

\bibitem[Siemiginowska \& Elvis(1997)]{siemiginowska97} Siemiginowska,
A., \& Elvis, M. 1997, \apj, 482, L9

\bibitem[Stoughton et al.(2002)]{stoughton02} Stoughton, C., Lupton,
R., et al. 2002, \aj, 123, 485

\bibitem[Terlevich et al.(1992)]{terlevich92} Terlevich, R.,
Tenorio-Tagle, G., Franco, J., \& Melnick, J. 1992, \mnras, 255, 713

\bibitem[Trevese et al.(1994)]{trevese94} Trevese, D., Kron, R. G.,
Majewski, S. R., Bershady, M. A., \& Koo, D. C. 1994, \apj, 433, 494

\bibitem[Trevese \& Vagnetti(2002)]{trevese02} Trevese, D., \&
Vagnetti, F. 2002, \apj, 564, 624

\bibitem[Vagnetti et al.(2003)]{vagnetti03} Vagnetti, F., Trevese, D., 
\& Nesci, R. 2003, \apj, 2003, June issue

\end{thebibliography}
\end{document}